\documentclass[reprint,aps,numerical,superscriptaddress,floatfix,nofootinbib]{revtex4-1}
\usepackage{amssymb}
\usepackage{amsthm}
\usepackage[figuresright]{rotating}
\usepackage{graphicx}
\usepackage{textcomp}
\usepackage[T1]{fontenc}
\usepackage[tight]{units}
\begin{document}
\title{Particle Discrimination in TeO$_{2}$ Bolometers using Light Detectors read out by Transition Edge Sensors}
\newcommand{\LNGS}{\affiliation{INFN - Laboratori Nazionali del Gran Sasso, Assergi (AQ) I-67010 - Italy}}
\newcommand{\sapienza}{\affiliation{Dipartimento di Fisica - Universit\`{a} di Roma La Sapienza, I-00185  Roma - Italy}}
\newcommand{\INFNROMA}{\affiliation{INFN - Sezione di Roma I, I-00185 Roma - Italy}}
\newcommand{\mpi}{\affiliation{Max-Planck-Institut f\"ur Physik, D-80805 M\"unchen - Germany}}
\newcommand{\AQ}{\affiliation{Dipartimento di Scienze Fisiche e Chimiche - Universit\`{a} degli studi dell'Aquila, I-67100 Coppito (AQ) - Italy}}
\newcommand{\GSSI}{\affiliation{Grans Sasso Science Institute, I-67100 L'Aquila - Italy}}
\author{K.~~Sch\"affner}
\email[corresponding author:]{karoline.schaeffner@lngs.infn.it}
\LNGS

\author{G.~Angloher}
  \mpi

\author{F.~Bellini}
 \sapienza 
 \INFNROMA

\author{N.~Casali}
 \LNGS
 \AQ

\author{F.~Ferroni}
 \sapienza 
 \INFNROMA

\author{D.~Hauff}
 \mpi

\author{S.S.~Nagorny}
 \GSSI

\author{L.~Pattavina}
 \LNGS

\author{F.~Petricca}
 \mpi

\author{S.~Pirro}
 \LNGS

\author{F.~Pr\"obst}
 \mpi

\author{F.~Reindl}
 \mpi

\author{W.~Seidel}
 \mpi

\author{R.~Strauss}
 \mpi
\begin{abstract}
An active discrimination of the dominant $\alpha$-background is the prerequisite for future neutrinoless double-beta decay experiments based on TeO$_{2}$ bolometers. We investigate such $\alpha$-particle rejection in cryogenic TeO$_{2}$ bolometers by the detection of Cherenkov light. For a setup consisting of a massive TeO$_{2}$ crystal (\unit[285]{g}) and a separate cryogenic light detector, both using transition edge sensors as temperature sensors operated at around \unit[10]{mK}, we obtain an event-by-event identification of e/$\gamma$- and $\alpha$-events. We find in the energy interval ranging from \unit[2400]{keV} to \unit[2800]{keV} and covering the Q-value of the neutrinoless double-beta decay of $^{130}$Te a separation of the means of the two populations of 3.7 times their width.  
\end{abstract}
\maketitle
\section{Introduction}
\label{intro}
The postulation of the neutrino in 1930 by W.~Pauli was followed by many decades of intensive experimental investigations, though still today important properties of this particle are unknown. Oscillation experiments have confirmed that the three families of neutrinos mix and that at least two of them have a finite mass. However, information on the absolute mass scale, the ordering of these masses, charge conjugation properties and lepton number conservation is still missing.\par
In case neutrinos are Majorana particles~\cite{Majorana}, which implies the presence of physics beyond the Standard Model of particle physics, an extremely rare process should be observable, namely the so-called Neutrinoless Double-Beta Decay (0$\nu$DBD) \cite{DBD}: in a 0$\nu$DBD the parent nucleus decays by the simultaneous emission of two beta-particles only. As no neutrinos are emitted, the full energy of the decay, the Q-value, is shared between the two electrons. The evidence of 0$\nu$DBD would prove that neutrinos are their own anti-particles and that lepton number is not conserved. Also, constraints would be set on the mass scale of the neutrinos.\par
Numerous experiments are searching for this process with the distinctive signature of a monochromatic line at the Q-value of the decay - the combined energy of the two simultaneously emitted electrons.\par
Low temperature bolometers are ideal detectors for such surveys: crystals can be grown with a variety of interesting DBD-emitters \cite{DBDemitter}, and multi-kg detectors~\cite{massive-tellurium} can be operated with excellent energy resolution (0.1-0.2\%) at the Q-value \cite{CUORE0}.\par
Up to now, low temperature bolometers searching for 0$\nu$DBD were mainly using TeO$_{2}$ crystals~\cite{CUORE0,CUORICINO}, which show very good mechanical and thermal properties~\cite{TeO2Crystals}, have a very large natural isotopic abundance of the candidate isotope $^{130}$Te (34.2\% \cite{130TeO}) and, most importantly, are produced at industrial scale.\par
At present, radioactive surface contamination is the key-issue~\cite{TeO2bck1,TeO2bck2} that may limit the sensitivity of tonne scale cryogenic bolometer experiments like CUORE~\cite{CUORE}: $\alpha$-particles can loose a fraction of their initial energy while passing through the bulk of the material surrounding the detector before interacting in the bolometer. These so-called degraded $\alpha$-particles show a flat energy spectrum ranging from the Q-value of the $\alpha$-decay (several MeV) down to threshold energy, thereby possibly creating background within the region of interest for 0$\nu$DBD \cite{TeO2bck3}.\par 
For next-generation bolometric 0$\nu$DBD experiments beyond CUORE~\cite{CUOREIHE}, the only way to eliminate this $\alpha$-background is to identify the interacting particles. In case of non-scintillating crystals like TeO$_{2}$ the discrimination could be obtained by measuring the Cherenkov light emitted by electrons as suggested in \cite{Tabarelli}. Alpha-particles of few MeV have kinetic energies below the threshold for the creation of Cherenkov light. Since the expected energy emitted in form of Cherenkov photons from electron interactions is $\mathcal{O}$(\unit[100]{eV})~\cite{Tabarelli}, light detectors with a low energy threshold and an excellent energy resolution should be employed.\par
The first published measurement on the detection of Cherenkov light in this context was carried out on a \unit[116]{g} TeO$_{2}$ crystal, demonstrating that $\alpha$-particles can be discriminated \cite{Cerenkov1}. Very recently an event-by-event discrimination was obtained with a very small TeO$_{2}$ bolometer based on a \unit[23]{g} crystal \cite{Willers}. Measurements on a \emph{massive} crystal (\unit[750]{g}) demonstrate that difficulties may substantially increase with crystal size and that light detectors with a threshold as low as some \unit[10]{eV} are required \cite{Cerenkov2}.\par
In this work we report the results from a measurement of the Cherenkov light emitted by a  \unit[285]{g} TeO$_{2}$ bolometer. The light absorber is read out by a Transition Edge Sensor (TES) of the same type as used in the CRESST dark matter search~\cite{CRESST}. With this set-up we demonstrate for the first time an effective event-by-event discrimination on a massive TeO$_{2}$ crystal.
\section{Experimental Set-up}
\label{sec:2}
\begin{figure}
\includegraphics[width=0.47\textwidth]{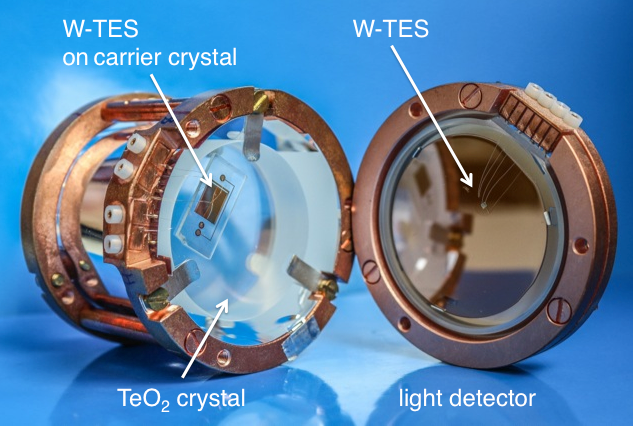}
\caption{On the left, the TeO$_{2}$ crystal surrounded by a reflective foil and mounted in its copper holder is shown. A small CdWO$_{4}$ crystal carrying the TES is attached onto the large TeO$_{2}$. The light detector shown on the right is mounted in its copper structure. The temperature sensors can be identified on the surfaces of both detectors.}
\label{fig:0}    
\end{figure}
The TeO$_{2}$ crystal used in this work is a cylinder (\unit[40]{mm} diameter and height, \unit[285]{g} in weight). The mantle surface of the crystal was mechanically roughened to increase the collection of Cherenkov light \cite{Danevich}. The two flat surfaces are of optical quality, polished with a \unit[1]{mm} chamfer on edges. The crystal is operated as a bolometer at about \unit[10]{mK} and read out by a W-TES. TESs are highly sensitive thermometers which allow to detect tiny temperature excursions $\mathcal{O}$($\mu$K) caused by particle interactions in the crystal, by measuring the change in the resistance of the TES.
In our case, the sensor consists of a thin tungsten film (\unit[200]{nm}, referred to as W-TES) stabilized in its transition between the normal conducting and the superconducting phase by a dedicated heater. The heater consists of a gold stitch bond (gold wire with diameter of \unit[25]{$\mu$m}), which is bonded to the tiny gold structure on the TES also used as thermal link. Since a direct evaporation onto the TeO$_{2}$ crystals did not succeed, a composite detector design was used instead \cite{CompositeDet}: a small \unit[(20x10x2)]{mm$^{3}$} cadmium tungstate carrier crystal (CdWO$_{4}$), equipped with the W-TES, was attached onto one of the flat surfaces of the TeO$_{2}$ by vacuum grease (DowCorning UHV). The reason for choosing a CdWO$_{4}$ crystal as carrier of the W-TES is discussed in section \ref{sec:3}.\par
One polished side of the TeO$_{2}$ crystal is facing a cryogenic light detector of CRESST-II type~\cite{CRESST}, which consists of a thin sapphire disc (thickness of \unit[460]{$\mu$m}) with a diameter of \unit[40]{mm}, equal to the diameter of the crystal. Since pure sapphire is optically transparent a \unit[1]{$\mu$m} thick layer of silicon is epitaxially grown onto the sapphire disc. Light detectors of this type we refer to as SOS (Silicon on Sapphire). Also the light absorber is read out by a W-TES, which in shape and dimension is optimized for the light detector and different to the TES design of the massive absorber crystal. Crystal and light detector are enclosed in a reflective housing (VM2002 Radiant Mirror Film). A photograph of the set-up is shown in Figure \ref{fig:0}.\par
\begin{figure}
\includegraphics[width=0.45\textwidth]{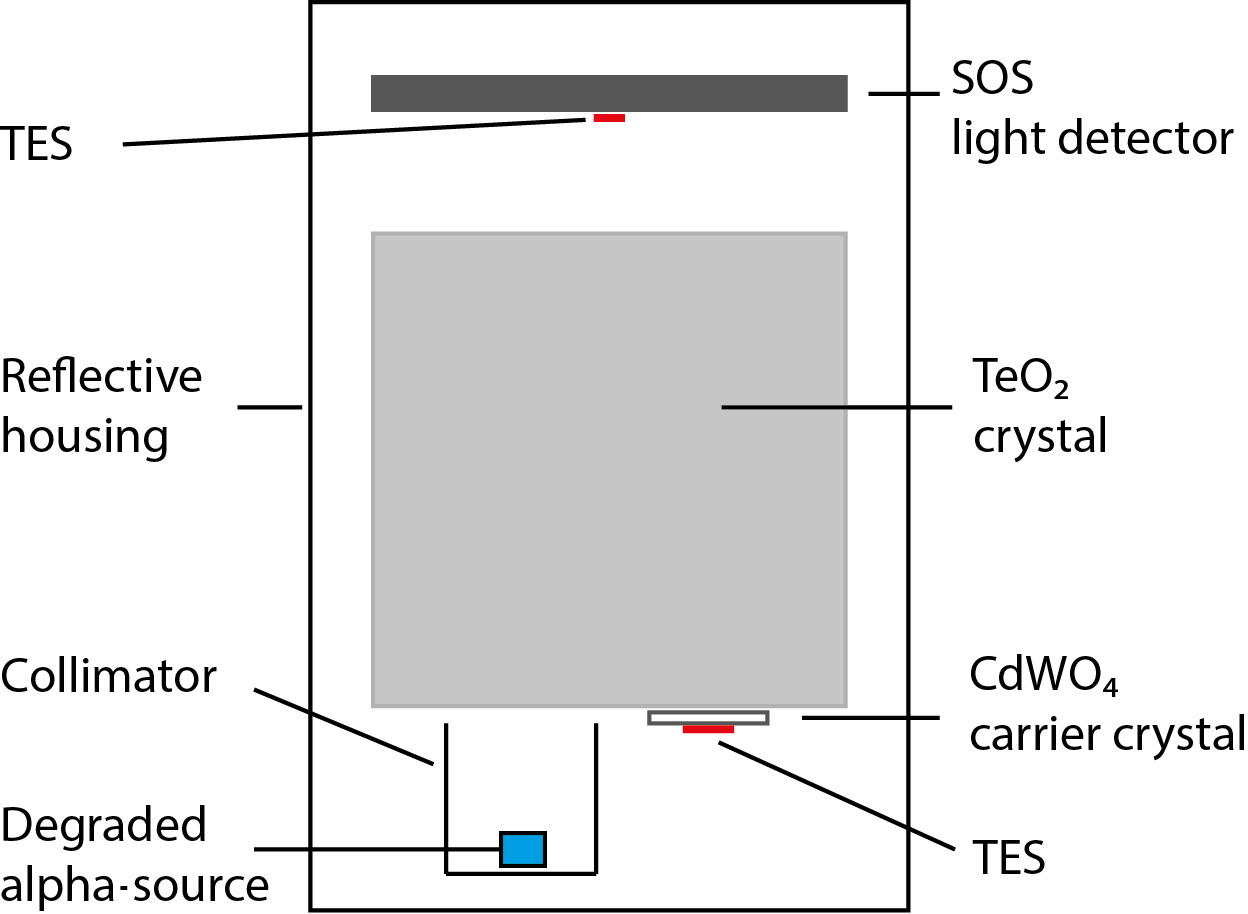}
\caption{Schematic of the experimental set-up (not drawn to scale).}
\label{fig:0b}    
\end{figure}

To study the discrimination capability of e/$\gamma$-events from $\alpha$-particles a source of degraded $\alpha$s of $^{238}$U is used. The source, as depicted in Figure \ref{fig:0b}, is arranged such to avoid any line of sight to the TES and to the surrounding.\par
The measurement was carried out in the test facility of the Max-Planck Institute for Physics. The facility is located at the Laboratori Nazionali del Gran Sasso (LNGS), a deep underground site (\unit[3650]{m.w.e.}) in central Italy and consists of a dilution refrigerator which is surrounded by about \unit[20]{cm} of low background lead to shield the environmental e/$\gamma$-radioactivity. A platform attached by a spring to the mixing chamber of the dilution unit mechanically decouples the detector from the cryogenic facility in order to reduce microphonic noise. The TESs are operated with a commercial dc-SQUID electronics (Applied Physics Systems). The hardware triggered signals are sampled in a \unit[409]{ms} window with a sampling rate of \unit[10]{kHz}. Both detectors are always read out simultaneously, independent of which channel triggered.\par
More detailed descriptions of the DAQ, the control of detector stability as well as the pulse height evaluation and energy calibration procedures are given in \cite{CRESST09,CRESST05}.

\section{Detector Performance}
\label{sec:3}
\begin{figure*}
{\includegraphics[width=0.49\textwidth]{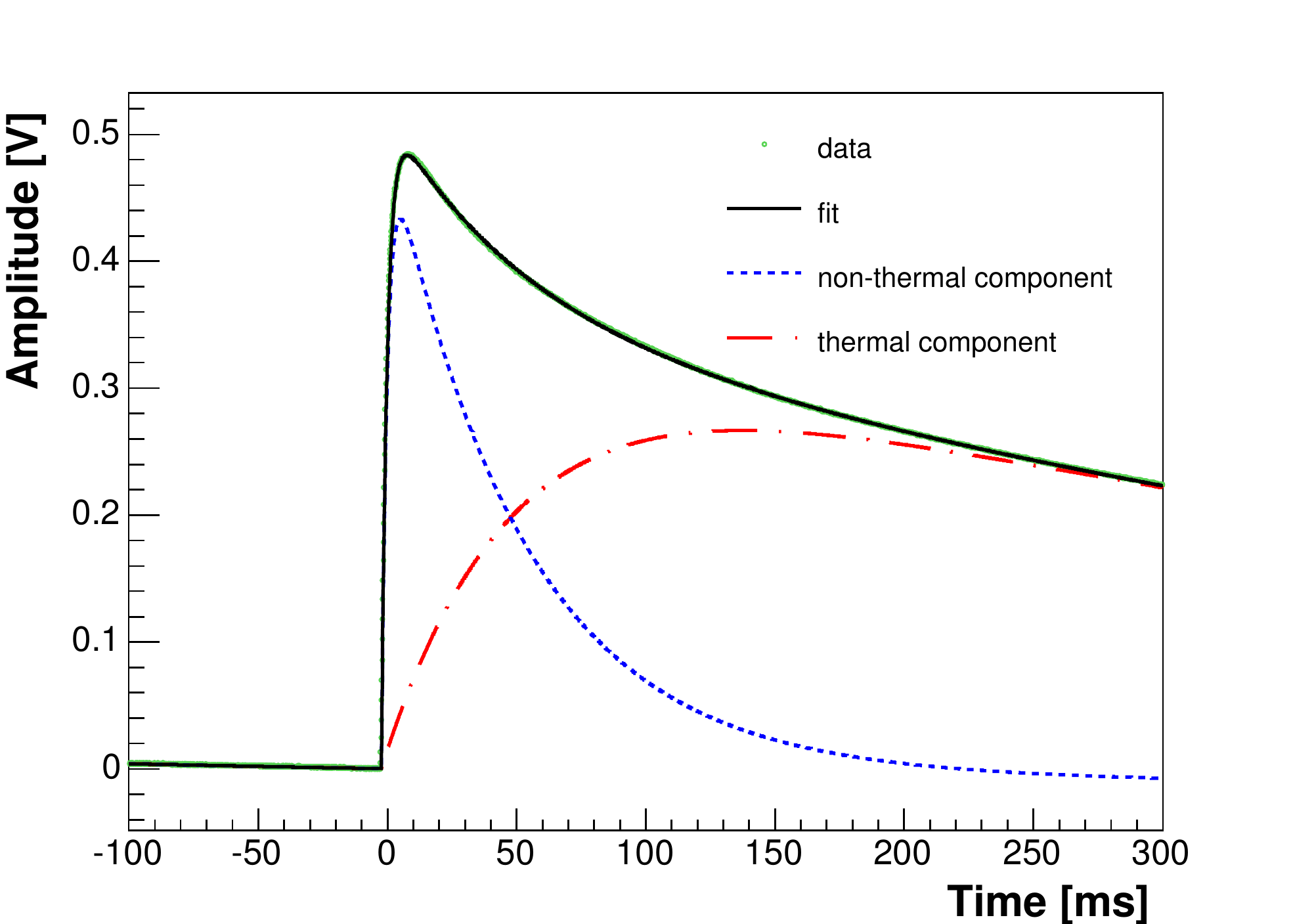}
  \includegraphics[width=0.49\textwidth]{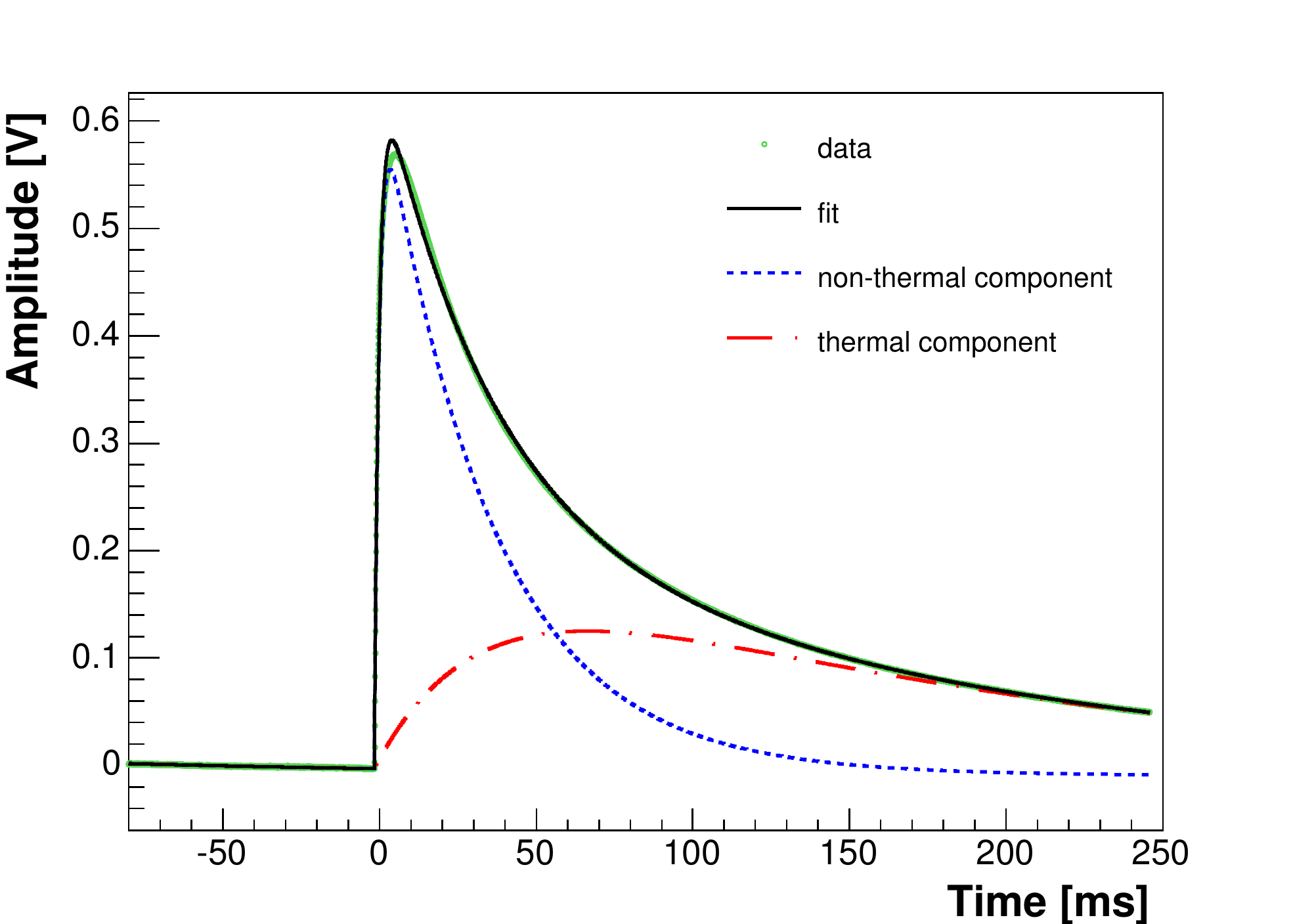}}
\caption{Left figure: Averaged pulse (green dots) created from \unit[2615]{keV} $\gamma$-events ($^{208}$Tl-line) in the TeO$_{2}$ crystal. Fitting this pulse with two exponentials using the model of cryogenic particle detectors \cite{Proebst} yields the non-thermal and thermal component, shown as dotted and dashed lines. Right figure: Averaged pulse (green dots) created from \unit[122]{keV} $\gamma$s ($^{57}$Co calibration) in a CdWO$_{4}$ crystal scintillator (\unit[40]{mm} in diameter and height) as well as its non-thermal (dotted line) and thermal component (dashed line). The black solid line referred to as "fit" represents the sum of the two components in both plots respectively.}
\label{fig:1}  
\end{figure*}
In this section we report on pulse-shape parameters and energy resolution of the TeO$_{2}$ bolometer and of its light detector.
\subsection{TeO$_{2}$ Crystal}
\begin{table}
\caption{Calculated phonon group velocities are given for acoustic transversal and acoustic longitudinal phonon modes in TeO$_{2}$, CaWO$_{4}$ and Al$_{2}$O$_{3}$. Since values for CdWO$_{4}$ are absent in literature, the values for CaWO$_{4}$ are listed instead. For each interface only the phonons travelling along the crystal axis normal to the interface are considered. Debye temperatures $\Theta_{D}$ of all crystals are listed.}
\label{tab:1}    
\begin{tabular}{lcccc}
\hline\noalign{\smallskip}
Material  & $v_{TA}$ &  $v_{LA}$ & $\Theta_{D}$ & Ref.\\
 &  [km/s]  &  [km/s]& [K] &  \\
\noalign{\smallskip}\hline\noalign{\smallskip}
 TeO$_{2}$ & 2.2 & 3.2 & 232 & \cite{Barucci}, \cite{Handbook}\\
 CaWO$_{4}$ & 2.45 & 4.76 & 335 & \cite{Senyshyn}, \cite{Head}\\
 Al$_{2}$O$_{3}$ & 6.1 & 11.2 & 1041 & \cite{Proebst}, \cite{Handbook}\\
\noalign{\smallskip}\hline
\end{tabular}
\end{table}
The measured signals in the TeO$_{2}$ bolometer are very small in comparison to other inorganic scintillating materials both read out by a W-TES of the type as used in this work \cite{Proebst, Schaeffner}. This behavior can be attributed to the acoustic mismatch between TeO$_{2}$ and the CdWO$_{4}$ crystal carrying the W-TES. A particle interaction in an inorganic anisotropic single crystal as, e.g.~TeO$_{2}$, creates high frequency phonons $\mathcal{O}$(THz) with energies of a few \unit[]{meV}. As these energies are much higher than thermal energies at the temperature of detector operation (about \unit[10]{mK}) corresponding to \unit[1]{$\mu$eV}, these phonons are called \textit{non-thermal phonons}.\par
This initial phonon population is not stable and very quickly decays to a distribution with a mean frequency of few \unit[100]{GHz} \cite{Proebst}. The fast decay is followed by a period of few milliseconds where the distribution of average phonon frequencies remains quasi constant. During this time the phonons spread ballistically over the volume of the crystal.\par
These phonons may get absorbed directly in the TES, thereby creating the fast so-called \textit{non-thermal part} of the detected signal. The TES relaxes back to equilibrium temperature via its thermal link to the heat bath (dominant process), but also via re-emission of thermal phonons into the crystal. Phonons that thermalize in the crystal by inelastic scattering (e.g.~on the surface or on crystal defects) also lead to an increase in the crystal's temperature  and cause the \textit{slow thermal component} of the detected signal. Typically, the fast component dominates the signal at very low operating temperatures, whereas the slow component is suppressed as a consequence of the weak coupling between electrons and phonons in the TES. Thus, being operated in the bolometric mode \cite{Proebst}, the pulse height of detected signals is determined by the flux of non-thermal phonons and by the thermal coupling of the TES to the heat bath.\par
The left plot of Figure \ref{fig:1} shows a fit to a pulse averaged over a set of $\mathcal{O}$(100) pulses from \unit[2615]{keV} $\gamma$-events ($^{208}$Tl-line) in the TeO$_{2}$ crystal. The pulse was fit with two exponentials using the model of cryogenic detectors \cite{Proebst}: the signal's non-thermal component (dotted line) and the thermal component (dashed line) are indicated. The sum of both components is shown as black solid line. For comparison, the right hand side of Figure \ref{fig:1} shows the result of such a fit to the events from \unit[122]{keV} $\gamma$s ($^{57}$Co calibration) in a CdWO$_{4}$ crystal (\unit[40]{mm} in diameter and height). For both measurements the same W-TES evaporated onto a small CdWO$_{4}$ carrier crystal was used. In case of the TeO$_{2}$ crystal the carrier was attached by conventional vacuum grease. In the case of CdWO$_{4}$, the carrier crystal was glued onto the big absorber using a low viscous epoxy resin (EpoTek 301-2).\par
For the pulses illustrated in Figure \ref{fig:1} the energy deposited in TeO$_{2}$ (left plot) is about 20 times higher than in the CdWO$_{4}$ crystal (right plot). However, the detected pulse amplitude is of same order. Especially, the contribution of the non-thermal component in the TeO$_{2}$ crystal is significantly reduced. Non-thermal phonons seem to thermalize before being detected in the TES, resulting in an overall small pulse amplitude with a long decay time.\par
The following explanation of the experimental observation is very plausible, however contributions from other processes cannot completely be excluded at this point. Phonons that have to be transmitted between two carefully bonded dissimilar crystalline media (crystal -$>$ interface -$>$ carrier crystal -$>$ metal film) will experience a boundary resistance. The reflection, refraction and mode conversion of phonons on crystalline interfaces is described by the acoustic impedance mismatch model \cite{Anderson}. In contrast, an amorphous interface as epoxy resin or grease allows low energy phonons more easily to pass the interface due to its rich energy spectrum, thus working as a low-pass filter \cite{Swatz}.\par
A first idea on the quality of the phonon transmission between solid-solid interfaces can be gained by comparing the Debye temperatures $T_{\Theta}$ of the bonded materials (see Table \ref{tab:1}). The more the $T_{\Theta}$ differs for two bond materials the lower the transmission probability.
The ratio of the speeds of sound $(v_{m1}/v_{m2})^{3}$ yields a qualitative estimate of the transmission probability from material 1 to material 2. A quantitative way of calculating the transmission and related parameters while including the anisotropy of a real crystal is illustrated in \cite{Weis}. However, such calculations cannot be performed here because the elastic constants and the lattice orientation are unknown for the materials of this work.\par 
The thin grease/epoxy layer used to attach the carrier crystal onto the large TeO$_{2}$ crystal does not affect the phonon propagation. This is proven by composite detectors made from CaWO$_{4}$ and CdWO$_{4}$ where no degradation of the phonon signal, in particular of the non-thermal component, is observed. Estimating the transmission probability for the material combination TeO$_{2}$ and CaWO$_{4}$ by using the simple ratio results in a transmission of about 30\%. Since CaWO$_{4}$ and CdWO$_{4}$ reveal similar crystal quality in sense of phonon propagation properties, a comparable value is expected for the material combination TeO$_{2}$ and CdWO$_{4}$.\par
Prior to the measurement reported in this article the same TeO$_{2}$ crystal was operated using a carrier crystal made out of sapphire (Al$_{2}$O$_{3}$). The pulse amplitude of \unit[2615]{keV} $\gamma$s was about six times smaller. The simple calculation for the phonon transmission results in about 3\% for the configuration TeO$_{2}$ and Al$_{2}$O$_{3}$ and is roughly consistent with the observed degradation in pulse amplitude in comparison to the combination TeO$_{2} $ crystal and CdWO$_{4}$ carrier crystal.\par
However, the acoustic mismatch between carrier crystal and the TeO$_{2}$ crystal is not sufficient to serve as the unique explanation for the observed degradation in the non-thermal signal component. In TeO$_{2}$ other processes which reduce the life time of the non-thermal phonon population due to inelastic processes in the crystal or on the crystals' surface may play a role. The mantle surface of the cylindrical TeO$_{2}$ crystal used in this work was mechanically roughened. Surface imperfections induced by a mechanical treatment showed a degradation of the detected signal amplitude, in other experimental works, using Si or sapphire substrates. Further studies are necessary to gain a better understanding of the observed small signal amplitudes in TeO$_{2}$ crystals read out by W-TES.\par
The energy resolution achieved at the \unit[2.6]{MeV} $^{208}$Tl $\gamma$-line is: $\sigma$=\unit[10.2]{keV}. In comparison, the energy resolution of a CUORE crystal (\unit[750]{g}) read out by a NTD-Ge with optimized thermal design and working conditions is \unit[2.2]{keV} $\sigma$ \cite{CUORE0}. An energy resolution of 0.1-0.2\%~at the Q-value energy is a prerequiste for a future 0$\nu$DBD experiment. Thus, thermistors to read out the signals of large bolometers are superior to transition edge sensors in the region of interest for 0$\nu$DBD (MeV-scale), given their large dynamic range of operation. Instead, W-TES are ideal sensors for the purpose of light detection and for cryogenic bolometers for dark matter searches since their performance is best at eV-keV scale.
\begin{figure}
\includegraphics[width=0.48\textwidth]{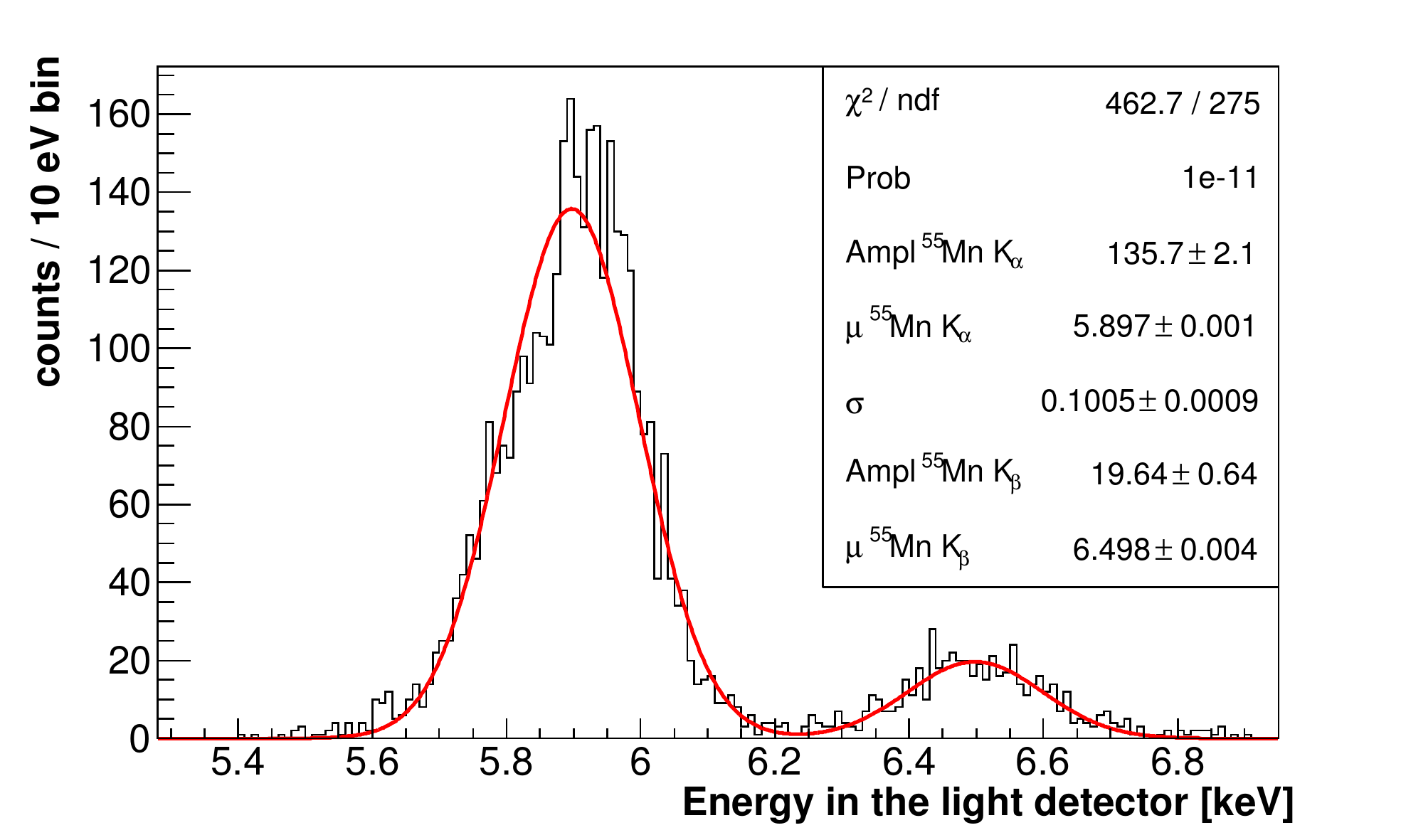}
\caption{Spectrum used to establish a direct energy calibration of the light detector. The $^{55}$Mn K$_{\alpha}$- and K$_{\beta}$-line at \unit[5.9]{keV} and \unit[6.5]{keV} show a resolution of \unit[(100.5 $\pm$ 0.9)]{eV} $\sigma$.}
\label{fig:2b}
\end{figure} 
\begin{figure*}
{\includegraphics[width=0.49\textwidth]{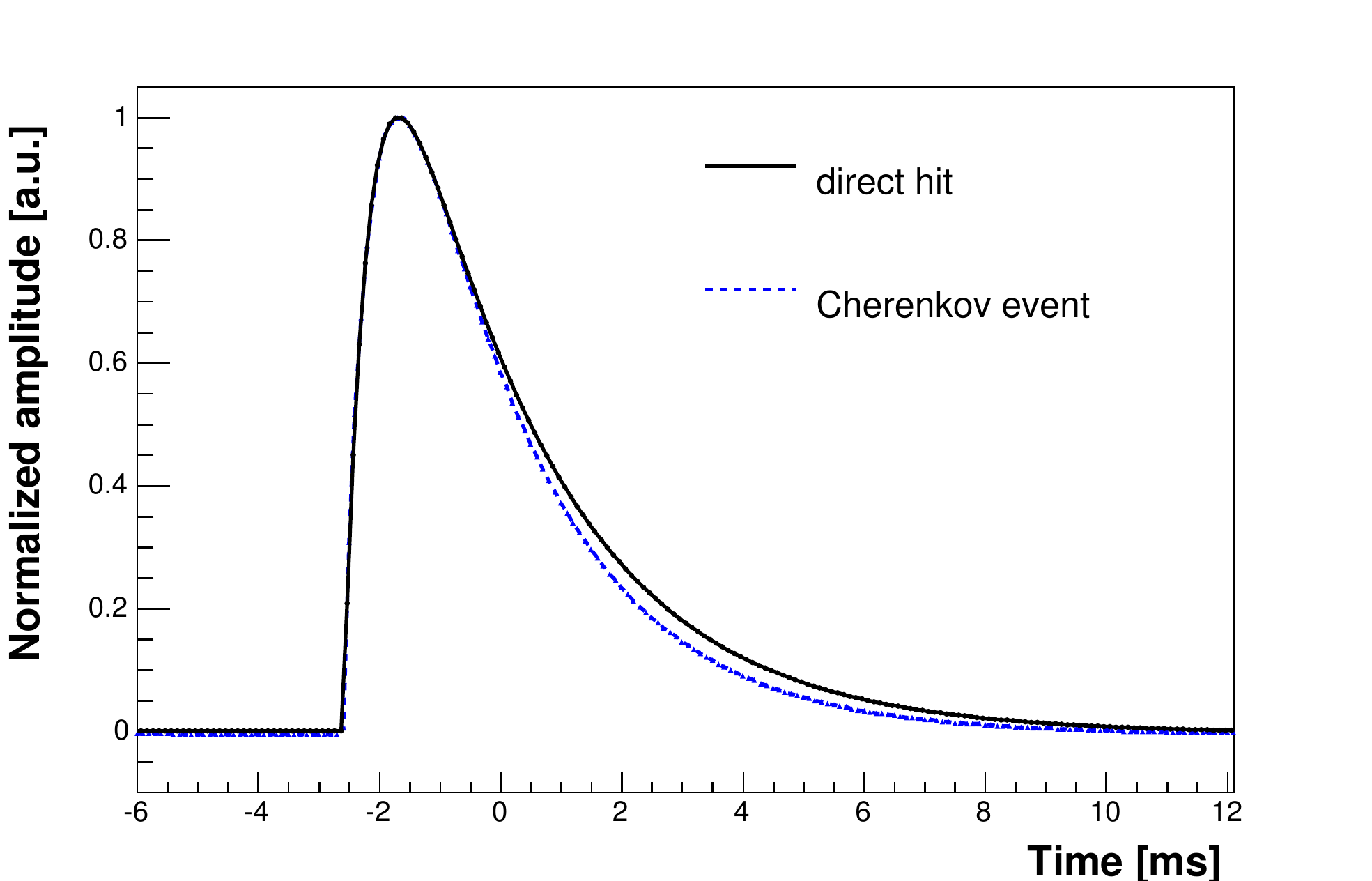}
  \includegraphics[width=0.49\textwidth]{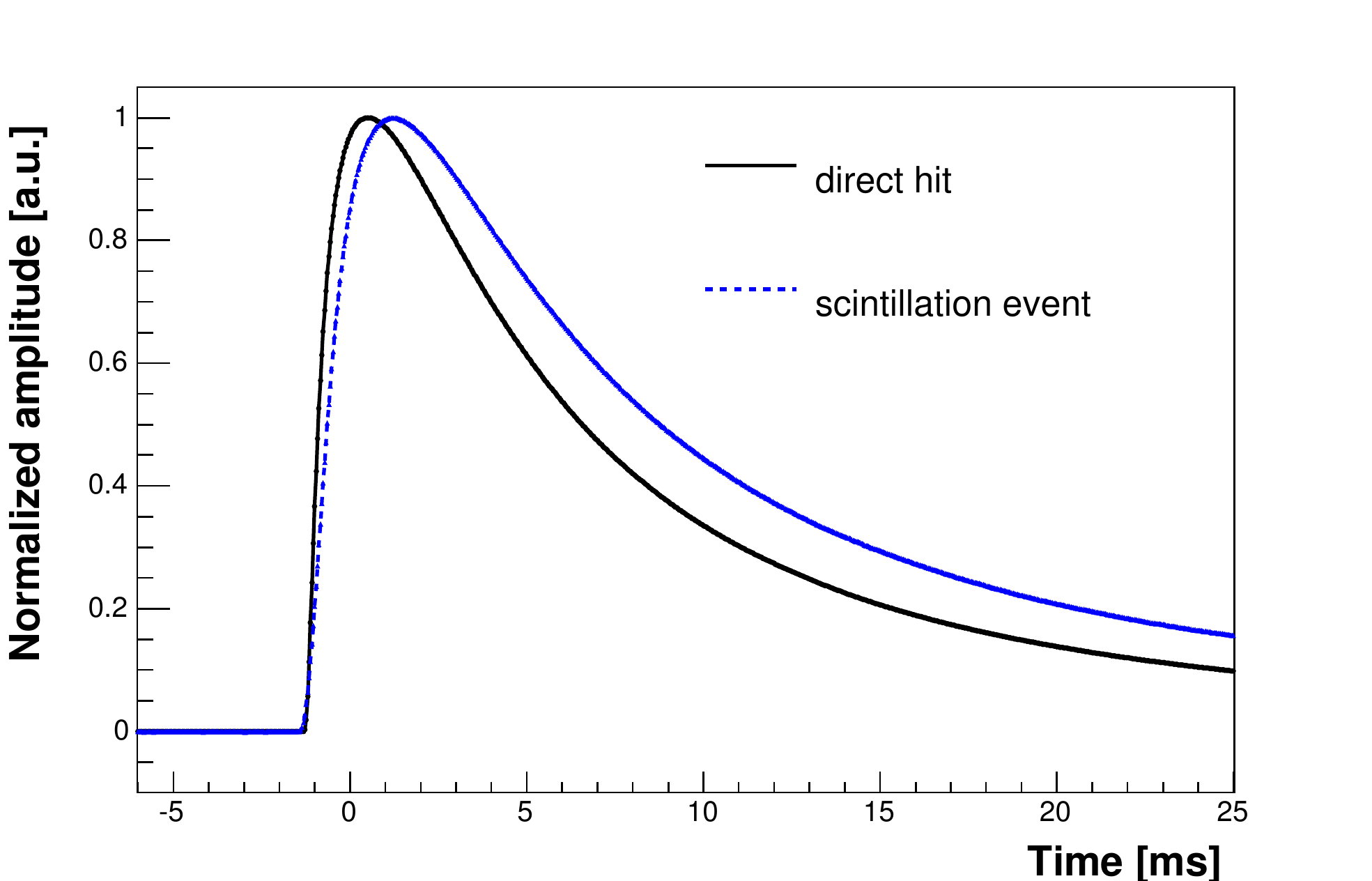}}
\caption{Left figure: The result of a fit to an averaged pulse (solid line) created from direct hit events in the light detector and from Cherenkov events (dashed line), both at an absolute energy of about \unit[125]{eV} (this work) is shown. Right figure: The result of a fit to an averaged pulse from direct hit events in a SOS light detector at \unit[5.9]{keV} ($^{55}$Mn K$_{\alpha}$-line) is displayed. Since this light detector is paired with a CaWO$_{4}$ crystal, the corresponding scintillation light event (same energy deposition in the light detector) from the CaWO$_{4}$ crystal is shown as a dashed line. As the scintillation process is slow in comparison to the production of the prompt Cherenkov light, the scintillation light event can be discriminated by pulse shape from a direct hit event in the light detector. A discrimination of a Cherenkov light event from a direct hit event in the light detector (left plot) is not feasible.}
\label{fig:2a}    
\end{figure*}
\subsection{Light Detector}
The SOS light detector has a transition temperature of about \unit[17.5]{mK}. An $^{55}$Fe X-ray source was used to allow for a direct energy calibration of the detector. The achieved energy resolution at the $^{55}$Mn K$_{\alpha}$- and K$_{\beta}$-line is about \unit[(100.5 $\pm$ 0.9)]{eV} $\sigma$ (see Figure \ref{fig:2b}), the RMS noise of the baseline is \unit[23.4]{eV} $\sigma$.\par
An investigation of the pulse shapes of the detected light signals shows that a pulse from a direct energy deposition in the light detector, referred to as a direct hit event, and a Cherenkov event reveal a very similar pulse shape for the same deposited energy (left plot in Figure \ref{fig:2a}). The detected light signal caused by an e/$\gamma$-interaction in the TeO$_{2}$ we refer to as Cherenkov event. The rise time of the non-thermal component of both signals is identical within errors; we find \unit[0.481$\pm$0.028]{ms} and \unit[0.488$\pm$0.008]{ms} for the direct hit event and the Cherenkov event, respectively.\par 
On the contrary, when using a typical inorganic scintillator made from CaWO$_{4}$, direct hit events and so-called scintillation light events, detected in the light detector as a consequence of a particle interaction in the CaWO$_{4}$ crystal, reveal a different pulse-shape as can be noted in the right plot of Figure \ref{fig:2a}: a direct hit event at \unit[5.9]{keV} in a SOS light detector (solid line) is shown together with an averaged pulse of a scintillation light event (dotted line) from the CaWO$_{4}$ crystal depositing the same amount of energy in the light detector.\footnote{The SOS light detector used with the CaWO$_{4}$ crystal is not the same as used in the TeO$_{2}$ measurement. However, they are from the same batch, both SOS-type and identical TES design.} The rise times are \unit[0.64]{ms} for the direct hit event and \unit[0.98]{ms} for the scintillation light event. This observation further confirms that the light detected from TeO$_{2}$ is Cherenkov light: the scintillation process is a slow process as it involves a sequence of stages, whereas the Cherenkov light is promptly produced by the charged particle. Therefore the pulse shape is similar to the one of particles (X-rays, electrons) directly absorbed in the light detector.
\section{Results}
\label{sec:4}
\begin{figure}
\includegraphics[width=0.47\textwidth]{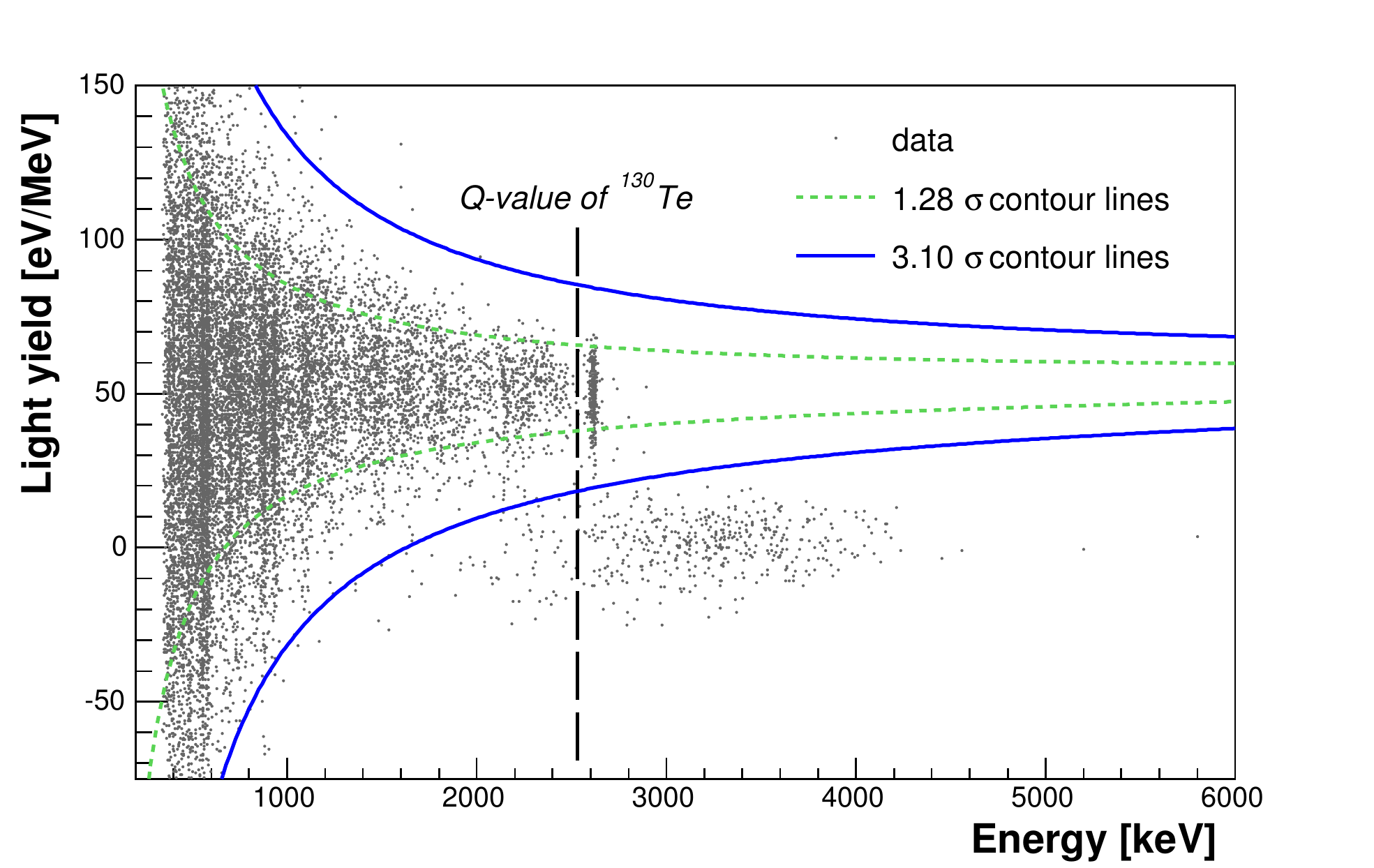}
\caption{Background data of the TeO$_{2}$ bolometer is shown in the light yield-energy plane. Light yield in this context refers to the direct energy detected in the light detector in eV per MeV of deposited energy in the TeO$_{2}$ crystal. Two distributions can be noted: a band due to e/$\gamma$-interactions as well as the less populated band at zero light yield due to $\alpha$-particles from a degraded $\alpha$-source. The bands which indicate the region expected for e/$\gamma$-events are shown in form of central probability bands. The dotted lines are $\pm$1.28 $\sigma$ contours whereas the solid lines are $\pm$3.1$\sigma$ contours, thus 99.8\% of all e/$\gamma$-events are expected to be contained within the two solid contour lines. The $\alpha$-particle distribution appears at a light yield of zero, separated from the populated e/$\gamma$-band. The dashed vertical line indicates the Q-value of $^{130}$Te of \unit[2530]{keV}. On average the Cherenkov light at the $^{208}$Tl-line results in about \unit[129]{eV} of detected light.}
\label{fig:4a}    
\end{figure}

The background data of the TeO$_{2}$ bolometer acquired in 0.67 days of live time in the light yield-energy plane is shown in Figure \ref{fig:4a}. The light yield (LY) in this context is defined as the direct energy detected in the light detector in eV per one MeV of deposited energy in the TeO$_{2}$ crystal. Two distributions can be observed. The highly populated band is due to e/$\gamma$-interactions, the less populated band at zero light yield is ascribed to $\alpha$-particle interactions from the degraded $^{238}$U $\alpha$-source.\par 
The test facility does not exhibit a low-background environment and several $\gamma$-lines are visible in the scatter plot (Figure \ref{fig:4a}). The dominant part of the e/$\gamma$-background comes from non-radiopure materials used for manufacturing the cryostat and the LHe-dewar located inside the Pb-shield. Furthermore, the Pb-shield around the LHe-dewar does not completely enclose the experiment. The highest observable $\gamma$-line from thallium appears at \unit[2615]{keV}. We find the energy of the detected Cherenkov light at this $^{208}$Tl-line to be \unit[128.9]{eV} ($\sigma$=\unit[32]{eV}).\par
The band ascribed to e/$\gamma$-events is shown in form of a central probability band. Two functions define this band: the \textit{mean value of the light yield} and the energy-dependent \textit{width} of the band around the mean value. The width is set by the finite energy resolution of the detectors. The following notation is used: the energy deposited in the crystal is $E$, the corresponding energy emitted in form of Cherenkov light is $L$. In this data-oriented model the mean of the light yield (LY) of the e/$\gamma$-event distribution is
\begin{equation}
LY_{e/\gamma}(E)= p_{0}E.
\label{eqn:mean}
\end{equation}
The value of $p_{0}$ is around \unit[50]{eV/MeV} since the light yield after absolute energy calibration of the light detector is about \unit[50]{eV} for \unit[1]{MeV} of deposited energy in the crystal.\par 
The width of the band is given by the energy resolution of the detectors. Due to high statistics, the energy dependence of the resolution can be extracted from a fit of the e/$\gamma$-band. The width of the e/$\gamma$-band can be well described by a Gaussian function with the width \cite{Strauss}
\begin{equation}
\sigma_{e/\gamma}=\sqrt{\sigma_{l}^{2}+\left(\frac{dL}{dE}\cdot\sigma_{TeO_{2}}\right)^{2}+S_{1}L}
\label{eqn:width}
\end{equation}
where $\sigma_{l}$ and $\sigma_{TeO_{2}}$ is the baseline noise of the light detector and the TeO$_{2}$ bolometer and $S_{1}$ accounts for statistical fluctuations in the number of detected photons (Poisson statistics). During the detector operation heater pulses are injected to the detector via the TES heater. The resolution of a heater pulse in the light detector and the TeO$_{2}$ bolometer directly yield the $\sigma_{l}$ and $\sigma_{TeO_{2}}$ parameters since heater pulses, in contrast to particle pulses, are not affected by photon statistics. The values for the detector module of this work are $\sigma_{l}$=\unit[261]{keV}, $\sigma_{TeO_{2}}$=\unit[9.8]{keV} and $S_{1}$=\unit[360]{keV}.\par

The band description is shown in Figure \ref{fig:4a}: the dotted contour lines are central $\pm$ 1.28 $\sigma$ boundary lines whereas the solid lines are $\pm$3.1$\sigma$ boundary lines, thus 99.8\% of all e/$\gamma$- events are expected to be within the two solid contour lines. The $\alpha$-particle distribution appears at a light yield of zero, well separated from the populated e/$\gamma$-band.\par
\begin{figure}
\includegraphics[width=0.47\textwidth]{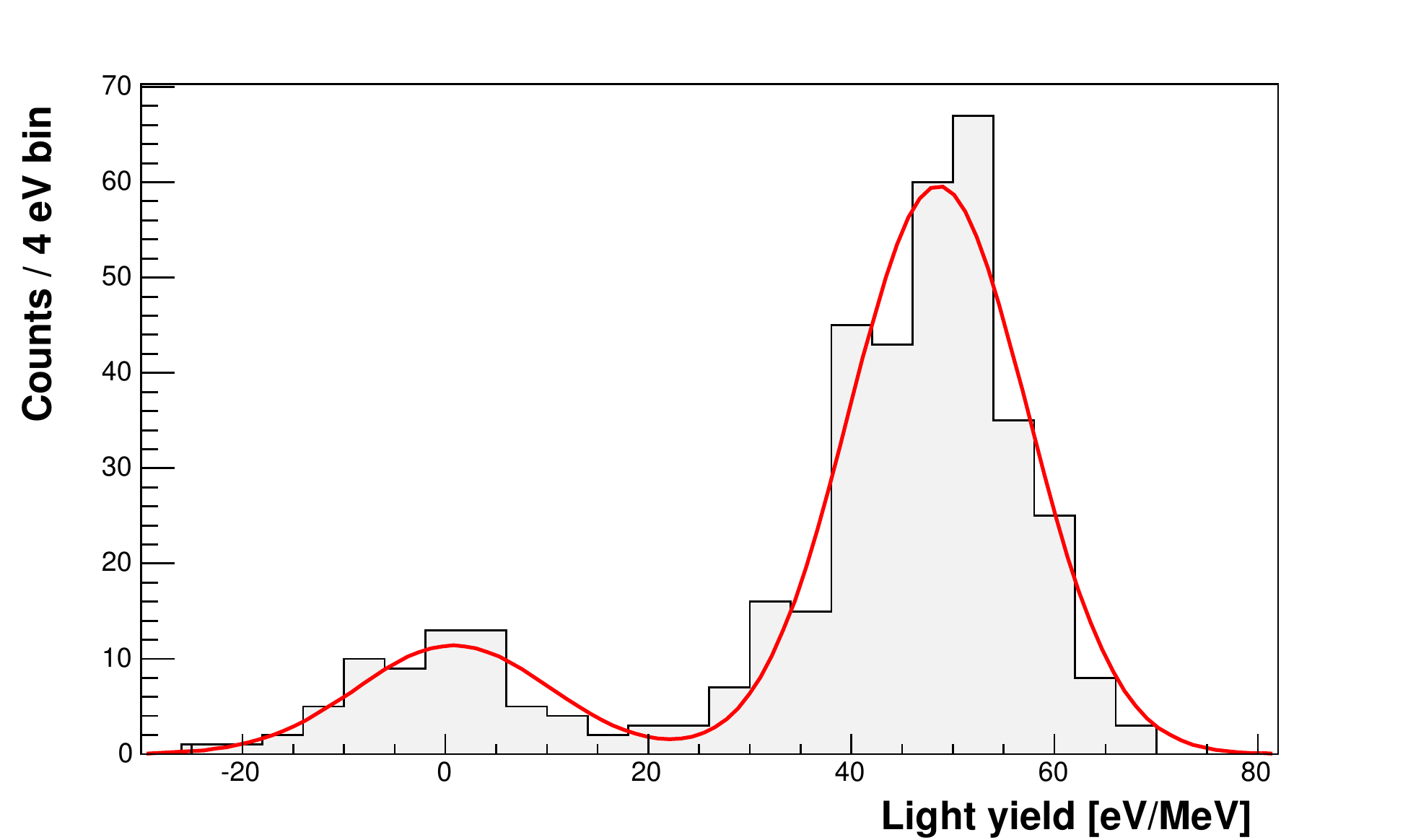}
\caption{A histogram of the light yield [eV/MeV] distribution is used to evaluate the discrimination power of e/$\gamma$-events from $\alpha$-particles in the region of interest. The distribution is fit with two Gaussians (all parameters free) and includes all events observed in the energy interval from \unit[2400]{keV} to \unit[2800]{keV}. We find a separation in this energy interval of 3.7 using Equation \ref{eqn:DP}.}
\label{fig:4b}     
\end{figure}

A projected view of the light yield in [eV/MeV] allows to visualize the achieved discrimination of $\alpha$-particles from e/$\gamma$-events; in Figure \ref{fig:4b} a histogram of all events observed in the energy interval from \unit[2400]{keV} to \unit[2800]{keV} is shown. Two Gaussian functions (all parameters free) are used to fit the two distributions (red solid line): $\alpha$-particles appear at \unit[(0.80$\pm$0.89)]{eV/MeV}, whereas for the distribution of e/$\gamma$-events the mean value is found to be \unit[(48.55$\pm$0.36)]{eV/MeV}.\par
The fit gives a resolution of $\sigma$=\unit[(8.74$\pm$0.27)]{eV/MeV} for the e/$\gamma$-peak and $\sigma$=\unit[(9.53$\pm$0.75)]{eV/MeV} for the $\alpha$-peak. Since the $\alpha$-events do not produce light, the resolution is expected to be slightly better in respect to the e/$\gamma$-events because of the missing Poisson statistics contribution. However, the $\sigma$s of both distributions are compatible due to the dominant contribution of the baseline noise and the lack of statistics (see Figure \ref{fig:4b}).\par
Following \cite{DP}, the discrimination power (DP) between two symmetric distributions can be characterized by comparing the difference between the mean values of the two distributions weighted by the square root of the quadratic sum of their widths
\begin{equation}
DP=\frac{\mu_{\beta / \gamma}-\mu_{\alpha}}{\sqrt{\sigma_{\beta / \gamma}^{2}+\sigma_{\alpha}^{2}}}
\label{eqn:DP}
\end{equation}
where $\mu_{\beta / \gamma}$ and  $\mu_{\alpha}$ are the mean values of the two distributions and $\sigma_{\beta / \gamma}$ and $\sigma_{\alpha}$ are their corresponding widths. Inserting the values obtained from the double-Gaussian fit displayed in Figure \ref{fig:4b} yields a value of DP equal to 3.7. Thus, we achieve the highest suppression up to date, in particular carried out on a massive TeO$_{2}$ bolometer (see Table \ref{tab:2}).\par 
Experiments using light detectors read out by NTD-Ge with the characteristics presented in \cite{Cerenkov1, Cerenkov2} so far do not arrive at a sensitivity level which allows for a discrimination on an event-by-event base. An improvement by a factor of four ($\sigma \approx$ \unit[20]{eV} (RMS)) on the performance of the thermistor would be needed for an effective $\alpha$-particle suppression.  A TES-based light detector combined with Neganov-Luke amplification technique \cite{Willers} showed an $\alpha$-suppression of 99\% while accepting 99.8\% of all e/$\gamma$-events at the $^{208}$Tl-line, but on a very small \unit[23]{g} TeO$_{2}$ crystal.\par
In Table \ref{tab:2} we summarize the properties and the performance of the before mentioned cryogenic light detectors used for the detection of Cherenkov light from TeO$_{2}$ bolometers.
\begin{table*}
\caption{Summary on the performance of various cryogenic light detectors used for the detection of Cherenkov light from TeO$_{2}$ bolometers.}
\label{tab:2}   
\begin{tabular}{l l l l l l l c}
\hline\noalign{\smallskip}
Material & Area [cm$^{2}$] & Thermometer & $\sigma$ & TeO$_{2}$ & Cherenkov light & DP & Ref.\\
absorber & light detector & light detector & [eV RMS]  & mass [g] & $@$ \unit[2.6]{MeV} in [eV] & & \\
\noalign{\smallskip}
\hline
\noalign{\smallskip}
 Ge  & 19.6 & NTD-Ge        & $\approx$ 72  & 750 & $\approx$ 100 & 1.5 & \cite{Cerenkov2} \\
 Ge  & 34.2 & NTD-Ge        & $\approx$ 97  & 117 & $\approx$ 195 & 1.4 & \cite{Cerenkov1} \\
 Si  & 4.0  & IrAu-TES + NL & $\approx$ 8   &  23 & $\approx$ 78  & 2.9 & \cite{Willers}\\
 SOS & 12.6 & W-TES         & $\approx$ 23  & 285 & $\approx$ 129 & 3.7 & this work \\
\noalign{\smallskip}\hline
\end{tabular}
\end{table*}
\section{Conclusion and Perspective}
In order to explore the inverted hierarchy region, future 0$\nu$DBD experiments based on TeO$_{2}$ bolometers have to adopt a particle discrimination method to reject the $\alpha$-background in the region of interest for 0$\nu$DBD. We demonstrated for the first time, using a massive TeO$_{2}$ bolometer (\unit[285]{g}), that the detection of the Cherenkov light by operating a TES-based cryogenic light detector allows to suppress the $\alpha$-background. We achieve a discrimination power of 3.7 for e/$\gamma$-events from $\alpha$-particles in the energy interval from \unit[2400]{keV} to \unit[2800]{keV} comprising the Q-value of $^{130}$Te.\par 
For the readout of the TeO$_{2}$ crystal a NTD-Ge thermistor is superior to a TES since such thermometers have shown to be able to reach a resolution of 0.1-0.2\%~at the Q-value of the double-beta decay of $^{130}$Te.\par
The light detector used in this work has an RMS of the baseline of about \unit[23]{eV} $\sigma$. Best performing TES-based light detectors operated in the CRESST-II dark matter search show $\sigma$=\unit[5]{eV} (RMS) \cite{Angloher14}. This improvement of a factor of about five would allow to enlarge the light absorber without loosing discrimination power. A light absorber with a larger area might be employed in order to detect the Cherenkov light from several massive TeO$_{2}$ crystals (\unit[750]{g} each), in a CUORE-like structure. To equip a next-generation CUORE-type experiment with 1-tonne of isotopic mass using enriched TeO$_{2}$ crystal with such large light absorbers would allow for a limited number of SQUID channels and would mitigate the challenge of manufacturing W-TES on a mass-production scale.
\section*{Acknowledgements}
This work was supported by the Italian Ministry of Research under the PRIN 2010ZXAZK9 2010-2011 grant. We want to thank the LNGS mechanical workshop and in particular E.~Tatananni, A.~Rotilio, A.~Corsi, and B.~Romualdi for continuous and constructive help in the overall set-up construction and to M.~Guetti for his constant technical support.
\bibliographystyle{h-physrev}
\bibliography{teo_arxiv.bib}
\end{document}